# The Future of Software Testing: AI-Powered Test Case Generation and Validation

Mohammad Baqar (baqar22@gmail.com) and Rajat Khanda(rajat.mnnit@gmail.com)

Abstract: Software testing is a crucial phase in the software development lifecycle (SDLC), ensuring that products meet necessary functional, performance, and quality benchmarks before release. Despite advancements in automation, traditional methods of generating and validating test cases still face significant challenges, including prolonged timelines, human error, incomplete test coverage, and high costs of manual intervention. These limitations often lead to delayed product launches and undetected defects that compromise software quality and user satisfaction.

The integration of AI in software testing has emerged as a transformative approach, addressing long-standing challenges in test case generation and validation. This paper explores AI-driven methods that leverage machine learning, natural language processing, and other advanced techniques to automate and enhance test case creation, optimize test coverage, and adapt to evolving software landscapes. Real-world examples illustrate how AI improves efficiency, accuracy, and scalability in testing workflows. However, the study acknowledges limitations, such as potential biases in AI algorithms, the need for substantial training data, and integration challenges with existing workflows. Additionally, areas for future research, including refining AI models to handle domain-specific testing scenarios and integrating AI testing with edge computing, are proposed. By presenting a comprehensive analysis of current tools, frameworks, and case studies, the paper advances understanding of AI's role in modern software testing and reports practical outcome patterns: faster triage, lower flaky-failure noise, stronger release confidence, and improved coverage quality when governance controls are explicit.

**Keywords**: AI in software testing, Automated test case generation, Test case validation, Machine learning in testing, Natural language processing (NLP) for testing, Test coverage optimization, Continuous integration and deployment (CI/CD), Predictive models in testing, Software quality assurance, Test automation, AI-driven testing tools, Self-healing test cases, Anomaly detection in testing, AI in software development, Test case optimization.

## 1. Introduction

Software testing remains central to the SDLC, yet conventional generation and validation practices struggle to match the speed, scale, and variability of modern delivery environments. This section positions AI as a practical mechanism for improving coverage quality, feedback velocity, and reliability.

### 1.1 Background

Software testing is a crucial component of the software development lifecycle (SDLC), ensuring that applications meet both functional and quality benchmarks prior to release. Its primary objective is to identify defects, validate functionality, and verify that software performs as expected across various scenarios. Testing acts as a safeguard, promoting the reliability, security, and performance of the product.

In the past, testing was primarily performed after the development phase, but modern methodologies like Agile and DevOps have integrated testing into each stage of development. Different testing strategies—such as unit, integration, system, and acceptance testing—are used to validate various components and layers of the application. Comprehensive testing helps uncover issues early, reducing the cost of corrections and enhancing the overall user experience. Without adequate testing, software risks performance failures, security breaches, and user dissatisfaction, which can harm the product and business reputation.

### 1.2 Challenges in Test cases Generation and validation

Traditional test case generation and validation methods face several critical challenges. One of the most pressing issues is the time-intensive nature of manually creating test cases, especially for large and complex systems. This process often requires significant effort from testers and can delay project timelines. Additionally, the reliance on human input introduces the risk of human error, where testers may overlook critical scenarios or fail to account for edge cases, leading to gaps in the testing process.

Another common issue is incomplete test coverage. It is difficult to ensure that every possible user interaction, code path, or functionality is thoroughly tested. As a result, untested areas of the code can harbor hidden defects that may not be discovered until later in production, potentially causing costly fixes. These challenges make traditional methods inefficient and less reliable in providing robust, comprehensive test validation.

### 1.3 Emergence of AI in testing

Artificial Intelligence (AI) is transforming the field of software testing by addressing many of the challenges associated with traditional methods. AI-powered testing tools are increasingly being integrated to automate the generation and validation of test cases, significantly reducing the time and manual effort involved. These tools can analyze code, learn from historical testing data, and automatically create test cases that cover a wide range of scenarios, including edge cases that might be overlooked by human testers.

AI also enhances test coverage by systematically identifying and prioritizing critical areas of the codebase that require testing, ensuring that all important functionalities are thoroughly examined. Additionally, AI algorithms can continuously adapt to changes in the code, enabling faster regression testing and reducing the need for manual updates to test scripts. By minimizing human error and automating repetitive tasks, AI improves testing accuracy, efficiency, and scalability. This integration of AI in testing is helping organizations achieve more reliable software with reduced effort and cost.

## 2. Literature Review

Research has progressed from conventional test design toward AI-augmented quality engineering. This section reviews that progression, explains why earlier foundations are stressed by rapid release cadences and distributed architectures, and identifies unresolved challenges that remain critical for dependable adoption.

### 2.1 Traditional Test Case Generation and Validation

Traditional test case generation and validation typically combine manual scenario design, scripted automation, and requirement-based planning. These approaches have historically provided traceability and control, especially in stable systems where behavior changes slowly and release windows are long. In such contexts, testers can map requirements to deterministic test suites and use predefined acceptance criteria to validate correctness with reasonable confidence.

However, the effectiveness of these methods declines as systems become more dynamic. Manual scenario derivation is time-intensive and susceptible to omission of edge conditions, while large scripted suites are costly to maintain when interfaces, dependencies, or business rules change frequently. Over time, organizations often accumulate brittle tests that are expensive to update and difficult to trust, which weakens feedback quality and slows delivery.

A further limitation is that traditional frameworks are not inherently risk-adaptive. They execute what is defined, but they do not readily infer where quality risk is increasing or where additional scenario exploration is needed. As a result, teams may spend disproportionate effort on low-impact checks while high-impact regressions remain under-tested. This imbalance has motivated the search for data-aware and adaptive testing methods.

### 2.2 AI in Software Testing

AI has emerged as a practical extension to conventional testing by introducing data-driven prioritization, adaptive scenario generation, and intelligent failure analysis. Instead of relying exclusively on static test plans, AI-enabled systems can learn from historical defects, runtime telemetry, and code-change patterns to recommend where testing effort should be concentrated.

Machine learning techniques contribute by modeling defect likelihood, ranking test execution order, and detecting anomalous behavior that may indicate hidden regressions. These capabilities are particularly useful in high-volume CI/CD environments, where exhaustive execution is often impractical and risk-based sequencing improves feedback speed without sacrificing quality confidence.

Natural language processing strengthens the connection between requirements and executable tests by extracting intent from user stories, specifications, and acceptance criteria. This reduces manual translation effort and improves alignment between business expectations and technical validation, especially when requirements evolve rapidly across iterations.

Large language models (LLMs) extend this capability by generating context-aware candidate scenarios, including boundary and negative paths that are frequently underrepresented in hand-authored suites. Their value is highest when integrated with deterministic controls such as schema constraints, policy checks, and human review, ensuring generated artifacts remain accurate, auditable, and relevant to project context.

### 2.3 Gaps in current Research

Despite strong progress, several research gaps continue to limit the reliability and scalability of AI-driven testing in production settings. One persistent challenge is interoperability with legacy systems, where outdated architectures, limited observability, and heterogeneous tooling complicate model integration and reduce the portability of AI-assisted workflows.

> Model transparency is another unresolved issue. Many high-performing models provide limited explainability, making it difficult for teams to justify release decisions, audit failure classifications, or trust automatically generated test recommendations. This concern is amplified in regulated domains that require traceable evidence for quality and compliance outcomes.
>
> Data quality remains a foundational dependency. AI systems trained on incomplete, noisy, or biased datasets can produce misleading priorities, fragile scenarios, or uneven coverage across user populations. Sustained model value therefore depends on disciplined data governance, continuous validation against ground truth, and explicit bias mitigation strategies throughout the testing lifecycle.
>
> Adaptability under continuous change is also an open problem. Modern software systems evolve through frequent commits, infrastructure updates, and shifting user behavior, and AI testing models must remain stable under these dynamics without constant manual recalibration. This requires research into robust online learning, drift detection, and low-overhead retraining methods.
>
> Human-AI collaboration frameworks need deeper empirical study as well. While AI can accelerate generation and triage, human testers remain essential for intent interpretation, exploratory reasoning, and ethical judgment. Effective operating models must define clear decision boundaries, escalation paths, and accountability mechanisms so automation strengthens rather than obscures quality governance.
>
> Scalability and performance considerations further shape adoption. As test volumes, repository sizes, and execution environments grow, AI-assisted workflows must maintain predictable latency and cost efficiency. Future work should evaluate architectural patterns that support high-throughput inference, selective execution, and resilient orchestration at enterprise scale.
>
> Security, privacy, and ethical risk management also require continued attention. AI-enabled testing pipelines may process sensitive logs, production-like datasets, and decision metadata; consequently, controls for data minimization, access governance, and safe model behavior are essential to prevent misuse or compliance violations.
>
> Finally, cost-effectiveness must be assessed with longitudinal evidence rather than short pilot outcomes. Organizations need robust methods to quantify return on investment across cycle-time reduction, escaped-defect prevention, maintenance effort, and operational risk reduction. Establishing this evidence base will be central to mainstream, sustainable adoption of AI-driven testing.
>
> Taken together, these gaps indicate that the future of AI in testing is not only a modeling problem but also a systems, governance, and organizational design problem. Progress will depend on integrating technical advances with strong engineering practices, transparent controls, and measurable quality outcomes.

Addressing these research challenges can materially improve the dependability of AI-assisted testing and enable broader adoption across complex software ecosystems, including safety-critical and compliance-sensitive domains.

### 2.4 Methodology

This paper uses a structured narrative-review methodology with tool-grounded evidence synthesis. Sources were selected from peer-reviewed software-engineering venues and high-quality technical documentation to capture both research validity and current operational practice. Selection prioritized empirical studies on AI-augmented test generation, prioritization, validation reliability, and LLM-based testing behavior in realistic development contexts [1], [23], [24], [25], [26], [27], [28], [29].

Evidence extraction followed four dimensions: (i) technical capability (generation, prioritization, self-healing, and validation), (ii) quality outcomes (defect yield, flakiness, and triage latency), (iii) governance properties (traceability, explainability, and policy control), and (iv) adoption constraints (integration cost, skill requirements, and scalability). To reduce bias, this review cross-compares research claims with deployment-oriented tooling patterns and CI/CD integration realities rather than relying on single-study conclusions.

Methodologically, the paper emphasizes socio-technical interpretation: AI testing effectiveness is evaluated not only by model output quality but also by its interaction with human review, organizational process maturity, and release governance. This framing supports conference-level reproducibility by making assumptions, evidence boundaries, and decision criteria explicit for future comparative studies.

## 3. Challenges in Traditional Test Case Generation and validation

Section 3 analyzes persistent challenges in traditional test case generation and validation, including manual effort intensity, incomplete scenario coverage, maintenance overhead, and human-error propagation. These limitations motivate the shift toward adaptive and intelligence-assisted quality engineering.

### 3.1 Manual Test Case Design

Manual test case design involves the painstaking process of creating test cases by hand, which can be both time-consuming and prone to errors. Testers must meticulously define each test scenario based on requirements, design detailed steps, and ensure that all relevant aspects of the application are covered. This process often involves a significant investment of time and effort, especially for complex or large-scale applications.

One of the major challenges of manual test case design is the risk of missing critical scenarios. Due to the sheer volume of possible interactions and edge cases, it's easy for testers to overlook certain conditions or paths in the application. Human oversight and limited experience can result in gaps in test coverage, which might leave important functionalities untested or bugs undiscovered. This can lead to defects slipping through into production, affecting software quality and user satisfaction.

Overall, while manual test case design is essential for understanding specific requirements and nuances, it often lacks the efficiency and comprehensiveness needed to keep pace with rapid development cycles and evolving software complexities.

### 3.2 Test Coverage Limitations

Ensuring comprehensive test coverage is a significant challenge, particularly in complex systems. Test coverage refers to the extent to which the test cases exercise different parts of the application, including code paths, functions, and user interactions. However, achieving thorough coverage presents several issues:

1. Complexity of Systems: In intricate systems with numerous components, interactions, and dependencies, it becomes difficult to create test cases that cover every possible scenario. The complexity increases exponentially with the addition of new features or changes, making it challenging to ensure all paths are tested.

2. Dynamic Nature of Software: As software evolves with updates, bug fixes, and new features, maintaining complete and up-to-date test coverage becomes a continuous task. Changes in the application can invalidate existing test cases or introduce new scenarios that need to be covered, complicating the testing process.

3. Edge Cases and Rare Scenarios: Identifying and creating test cases for edge cases or rare scenarios can be particularly challenging. These conditions might not be evident during initial test design, and their exclusion can result in incomplete coverage and undetected defects.

4. Resource Constraints: Comprehensive testing requires significant resources, including time, expertise, and computational power. In resource-constrained environments, it may not be feasible to design and execute enough test cases to achieve full coverage.

5. Human Limitations: Test designers may inadvertently overlook certain areas due to cognitive limitations or lack of experience, leading to gaps in coverage. Manual test case creation is inherently limited by the knowledge and attention of the testers.

6. Automation Challenges: While automated testing tools can enhance coverage, they are not immune to limitations. Automated tests need to be carefully designed and maintained to ensure they cover all relevant scenarios, and they may struggle with highly dynamic or complex user interactions.

Addressing these limitations requires a combination of advanced testing strategies, including AI-driven tools, risk-based testing approaches, and continuous integration practices to improve test coverage and ensure robust software quality.

### 3.3 Maintenance of Test Cases

Maintaining test cases in the face of frequent software changes is a significant challenge in software testing. As applications evolve with updates, bug fixes, and new features, ensuring that test cases remain accurate and relevant requires continuous effort. Here are some key difficulties associated with test case maintenance:

1. Frequent Updates: Software is often updated with new features, enhancements, or changes to existing functionality. Each update may necessitate changes to existing test cases or the creation of new ones to reflect the updated functionality. This ongoing need to adjust test cases can be labor-intensive and prone to oversight.

2. Test Case Obsolescence: Test cases can become obsolete or irrelevant as the software evolves. Outdated test cases that no longer match the current version of the software can lead to false results or missed defects. Keeping test cases synchronized with the latest version of the software requires regular reviews and updates.

3. Complexity of Changes: For complex applications with numerous interconnected components, even minor changes can have widespread effects. This complexity makes it challenging to determine which test cases need to be updated, and ensuring comprehensive coverage after changes can be cumbersome.

4. Resource Constraints: Maintaining and updating test cases demands significant resources, including time, expertise, and tools. In environments with limited resources, the task of updating and managing test cases can be overwhelming, leading to gaps in testing.

5. Automated Test Maintenance: While automated tests can enhance efficiency, they also require careful maintenance. Automated test scripts must be adjusted or rewritten to align with changes in the application, and automated test maintenance can become complex and require ongoing attention.

6. Coordination with Development: Effective test case maintenance requires close coordination with development teams to understand changes and their impact on testing. Communication gaps between developers and testers can result in discrepancies between the software's current state and the test cases.

Addressing these maintenance challenges involves implementing robust test management practices, utilizing AI-driven tools for dynamic test updates, and adopting agile methodologies to align testing with continuous development cycles.

### 3.4 Human Error

Human error plays a significant role in both the generation and validation of test cases, impacting the overall effectiveness and reliability of the testing process. The following are key aspects of how human oversight affects testing:

1. Inaccurate Test Case Design: When designing test cases manually, testers may overlook critical scenarios or fail to anticipate edge cases. This oversight can lead to incomplete test coverage and missed defects, compromising the quality and reliability of the software.

2. Inconsistencies in Test Execution: Variability in how test cases are executed by different testers can introduce inconsistencies. Testers may interpret test steps differently or make errors in executing the tests, leading to inconsistent results and difficulties in reproducing issues.

3.Errors in Test Data Management: Test cases often rely on specific test data to simulate various conditions. Human errors in creating or managing this data can result in incorrect test outcomes, affecting the accuracy of the validation process.

4. Maintenance Challenges: As software evolves, updating test cases to reflect changes can be prone to human error. Testers may inadvertently fail to update or modify test cases correctly, leading to outdated or irrelevant tests that do not align with the current version of the software.

5. Manual Review Limitations: In manual testing, the process of reviewing test results and identifying defects is susceptible to human error. Testers might miss subtle issues or misinterpret results, leading to incomplete defect identification and unresolved problems.

6. Communication Gaps: Miscommunication between development and testing teams can lead to misunderstandings about requirements or changes. This gap can result in test cases that do not accurately reflect the intended functionality or impact of changes.

To mitigate the impact of human error, organizations can employ strategies such as automated testing tools, rigorous review processes, and clear communication channels. Additionally, incorporating AI-driven tools can help reduce reliance on manual processes and enhance the accuracy and consistency of test case generation and validation.

## 4. AI-Driven Test Case Generation

Section 4 presents a research-oriented view of AI-driven test case generation, focusing on how modern models convert heterogeneous engineering artifacts into executable and high-value test suites. Rather than treating generation as a one-step prompt task, current approaches implement a multi-stage pipeline that combines requirements understanding, code-structure analysis, historical defect learning, and risk-aware prioritization. This pipeline design is essential for maintaining both scale and reliability in fast-moving software programs. From a methodological standpoint, recent work emphasizes hybrid architectures in which probabilistic generation is bounded by deterministic constraints, enabling teams to increase scenario breadth without sacrificing reproducibility or governance.

### 4.1 Machine Learning for Test Case Design

Machine learning contributes to test case design by learning defect-relevant patterns from source code, change history, execution logs, and prior test outcomes. In mature implementations, feature extraction spans static properties (complexity, dependency centrality, ownership churn) and dynamic properties (runtime error signatures, latency anomalies, flaky behavior). These features are then used to rank risk and guide scenario generation toward areas with the highest expected defect density. Empirically, the most useful models are those that optimize for defect yield per execution budget rather than raw test count, because this objective aligns with practical CI/CD constraints.

A key methodological step is structural code representation. Graph-based and sequence-based models can capture call dependencies, control-flow transitions, and module coupling, allowing generation systems to identify where boundary conditions and integration failures are most likely. This improves over purely requirement-driven generation, which often under-represents non-functional and cross-component failure modes. In large codebases, representation learning also helps detect latent coupling between modules, which is frequently where cascading regressions originate.

Requirements analysis remains central, but modern systems no longer parse requirements in isolation. They align requirement text with API contracts, user stories, acceptance criteria, and domain vocabulary. This cross-artifact alignment reduces semantic drift between intended behavior and generated tests, particularly in domains where business rules evolve rapidly and documentation quality is uneven. Teams that operationalize this alignment typically maintain requirement-to-test trace matrices, enabling faster impact analysis when scope changes late in the release cycle.

Historical defect mining further improves precision. By learning recurring failure motifs, models can prioritize scenarios resembling known incident trajectories, such as authorization misconfigurations, state-synchronization errors, or concurrency regressions. This historical grounding is one of the most practical ways to improve early defect discovery in CI/CD pipelines. To avoid overfitting to legacy failure classes, strong implementations periodically rebalance training windows and incorporate fresh incident evidence from production telemetry.

Adaptive learning is another research priority. As codebases evolve, model performance can degrade due to concept drift. Effective systems therefore include periodic retraining, drift detection, and feedback loops from failed tests and post-release incidents, ensuring generated suites remain relevant under continuous delivery conditions. A useful operational pattern is to define explicit drift thresholds (for example, prediction calibration loss or defect-yield drop) that trigger model retraining and regression benchmarking.

Risk-based generation should be interpreted as a resource-allocation strategy rather than a shortcut. High-impact modules, externally exposed interfaces, and frequently changed components require deeper scenario exploration, while low-risk areas can be validated with lighter test depth. This selective intensity improves throughput without weakening release confidence. In practice, organizations often encode this as policy tiers so that critical domains (payments, identity, safety logic) automatically receive stricter test-depth requirements.

Despite strong results, ML-based generation has known limitations, including explainability constraints, bias sensitivity, and overfitting to historical failure distributions. A defensible research posture therefore combines probabilistic generation with deterministic constraints, policy checks, and human review checkpoints. Additional safeguards such as adversarial test probes and counterfactual analysis can help teams verify that model recommendations are stable under plausible requirement and input variations.

Taken together, machine learning transforms test design from static enumeration into adaptive risk modeling. The strongest outcomes occur when generated artifacts are continuously evaluated against defect yield, flakiness rate, maintenance burden, and escaped-defect trends rather than raw test-volume growth [5]. For research-grade reporting, these outcomes should be tracked longitudinally across releases, with ablation studies that isolate the contribution of generation, prioritization, and optimization components.

### 4.2 Natural Language Processing (NLP) in Test Case Generation

NLP methods address one of the oldest gaps in software quality engineering: the distance between human requirements and executable verification logic. In modern pipelines, NLP models extract intents, entities, conditions, and constraints from specifications and user stories, then translate them into structured test objectives that can be mapped to frameworks and environments. The research challenge is not just extraction accuracy, but semantic faithfulness: preserving original business intent while converting natural language into precise, machine-actionable scenarios.

Advanced requirement parsing goes beyond keyword extraction by modeling semantic roles and dependency relations, which helps distinguish mandatory behavior from optional or contextual behavior. This distinction is important when generating tests for edge cases, validation rules, and error handling flows that are often ambiguously specified in natural language. Domain ontologies and custom vocabularies further improve precision by resolving project-specific terminology that generic models may misinterpret.

NLP can also detect specification inconsistencies before implementation defects appear. Contradictions across requirements, missing preconditions, and untestable acceptance statements can be flagged during planning, reducing downstream rework and improving traceability between product intent and verification scope. This early detection effect is especially valuable in distributed teams where product and engineering artifacts are authored by different stakeholders over asynchronous cycles.

When combined with LLM-based generation, NLP pipelines can produce richer scenario variants, including negative, boundary, and adversarial paths. However, reliability requires constrained generation strategies such as schema-guided prompts, domain dictionaries, and template validation, because unconstrained output can produce plausible but non-executable tests. A robust pattern is two-pass generation: first produce candidate scenarios, then perform deterministic conformance checks before execution admission.

Traceability is a critical research concern in this area. High-quality systems maintain explicit links among requirements, generated scenarios, execution evidence, and defect reports. These links make it possible to audit coverage quality and justify release decisions in environments with strict governance or compliance requirements. They also support post-incident analysis by showing which requirement assumptions were validated, missed, or incorrectly interpreted.

Continuous requirement evolution creates an additional challenge. NLP-assisted systems must monitor requirement deltas and trigger targeted regeneration of affected tests, rather than reauthoring entire suites. This selective regeneration reduces maintenance effort and improves change responsiveness. Version-aware diffing of requirement documents can further reduce noise by isolating semantic changes from superficial wording updates.

NLP also improves collaboration quality across technical and non-technical stakeholders by converting informal business language into testable constructs while preserving domain intent. In practice, this reduces ambiguity during sprint planning and clarifies acceptance boundaries between product, engineering, and QA teams. It also enables clearer negotiation of test scope when delivery timelines and risk tolerance need to be balanced explicitly.

Overall, NLP-based generation is most effective when coupled with deterministic validation and human oversight. The goal is not to automate interpretation blindly, but to improve consistency, traceability, and speed in requirement-to-test transformation. Future research should compare extraction and generation quality across domains, since domain-specific language strongly affects model behavior and test utility.

### 4.3 Automated Test Optimization

Automated optimization is the control layer that keeps generated suites executable, maintainable, and cost-efficient at scale. Without optimization, AI-driven generation can inflate suite size, amplify redundancy, and increase flaky behavior, ultimately degrading confidence rather than improving it. Optimization therefore functions as a quality governor that converts raw generation output into a reliable decision signal for release management.

Redundancy management typically combines similarity clustering, impact analysis, and dominance filtering to remove low-value duplicates while preserving fault-detection diversity. This is particularly important in large repositories where repeated scenario patterns can consume significant execution time without improving defect yield. Advanced implementations weight redundancy decisions by historical mutation sensitivity so pruning does not silently remove high-value assertions.

Prioritization models use risk features, business criticality, and change impact to reorder execution so that high-consequence regressions are evaluated first. In continuous delivery, this ordering materially improves decision latency by surfacing release-blocking failures earlier in the pipeline. Prioritization quality should be measured by time-to-critical-defect and not only by overall pass-rate statistics.

Data-driven optimization expands scenario depth without linearly increasing manual authoring. Parameterized inputs, boundary sets, and generated data profiles allow a smaller set of templates to cover broader behavioral space, especially for API and stateful workflow testing. The strongest designs also include data-validity constraints to prevent unrealistic synthetic inputs from inflating apparent coverage quality.

Refactoring assistance and flakiness analytics are increasingly integrated into optimization engines. By analyzing unstable assertions, timing-sensitive steps, and environment-dependent behavior, these systems can recommend structural improvements that reduce false failures and maintenance churn. Over time, flakiness reduction has a compounding effect by restoring trust in automated gates and lowering triage overhead across teams.

Change-aware regeneration is another major advance. Instead of re-running and re-authoring everything, optimization systems map code deltas to impacted test regions and regenerate only where coverage confidence drops. This targeted approach improves both efficiency and reproducibility. It also enables finer-grained ownership, since teams can attribute regeneration outcomes to specific services and change sets.

Self-healing behavior can reduce operational friction, but it requires guardrails. Automatically fixing selectors or locators is useful for minor UI shifts; however, behavior-level changes should trigger review rather than silent acceptance, otherwise defect masking risk increases. A practical governance rule is to permit autonomous fixes for non-semantic UI drift while requiring approval for any change that could alter business behavior validation.

Coverage optimization should also move beyond line coverage alone. Robust programs combine structural coverage, mutation score, and incident-relevance metrics to evaluate whether tests are meaningfully sensitive to real faults rather than simply executing large volumes of code. Multi-metric reporting is essential because single metrics are easy to optimize without improving actual defect prevention.

Cost-aware scheduling completes the optimization loop. Execution resources are finite, so high-maturity pipelines allocate runtime budgets dynamically across unit, integration, end-to-end, performance, and security checks according to release risk and service criticality. This allows teams to maintain quality assurance under tight delivery windows without uniformly increasing infrastructure spend.

In research terms, automated optimization is where AI-driven generation becomes operationally viable. It converts raw test abundance into a calibrated quality signal that is faster to execute, easier to maintain, and more aligned with business risk. Future work should investigate optimization policies that jointly minimize cost, latency, and escaped-defect probability under realistic enterprise constraints.

### 4.4 Tools and Frameworks (Modern Ecosystem)

The current tool landscape reflects a shift from monolithic test automation toward composable quality stacks. Organizations typically combine specialized tools across UI, API, integration, performance, security, and observability rather than relying on a single platform for all validation tasks. This composability is important because modern software quality problems rarely live in a single layer.

For web and end-to-end layers, Playwright and Cypress are now widely adopted because they provide robust browser automation, strong debugging ergonomics, and stable execution in CI environments. Selenium 4 remains relevant for cross-browser compatibility and legacy migration pathways in enterprise systems. Tool choice is increasingly driven by maintainability and failure diagnosability rather than just script execution speed [11], [16], [17].

At service and integration layers, Postman/Newman and Pact support API contract confidence, while Testcontainers enables realistic dependency orchestration in ephemeral test environments. This combination significantly reduces false confidence caused by mocked-only test strategies. It also improves defect reproducibility by aligning test-time dependencies more closely with production behavior [18], [19].

Mobile and cross-device validation increasingly relies on Appium 2 with cloud device infrastructure to widen coverage across operating-system and hardware variants. This has become essential for teams targeting heterogeneous mobile ecosystems. In practice, organizations often pair this with risk-based device matrices to balance coverage depth with execution cost [20].

For non-functional validation, k6 and Gatling are commonly used to integrate performance and resilience checks into delivery pipelines, enabling teams to detect capacity regressions before production exposure. The most mature pipelines treat these checks as first-class release gates instead of post-deployment diagnostics [21], [22].

Observability-driven validation has become a defining capability, with OpenTelemetry enabling trace, metric, and log correlation during test execution. This allows teams to verify distributed behavior, not only external outputs, and improves root-cause resolution for intermittent failures. It also strengthens causal analysis by linking failed assertions to specific service paths and infrastructure conditions [10].

Security and governance layers are increasingly integrated through tools such as CodeQL, Semgrep, and OWASP ZAP, which extend test pipelines into static and dynamic risk detection. This integration supports shift-left quality gates that combine functional correctness with security posture. As regulatory pressure grows, these controls are becoming essential rather than optional for enterprise release assurance [12], [13], [14].

Mutation-oriented adequacy tools such as Stryker provide an additional signal on assertion strength by testing whether suites can detect controlled behavioral perturbations. This helps distinguish genuinely effective tests from high-volume but low-sensitivity suites. Mutation trends over time also provide a useful leading indicator of test-suite health and maintenance quality [15].

In practice, these tools are orchestrated through CI/CD platforms with policy gates for release decisions. The key insight is that tool effectiveness depends less on individual features and more on how well the stack is integrated with risk models, feedback loops, and governance controls. Integration architecture therefore deserves as much design attention as tool selection itself [6], [7].

Consequently, modern frameworks should be evaluated not only by automation speed, but by their ability to improve defect detection quality, traceability, maintainability, and release confidence under continuous change. Comparative evaluations should include operational metrics such as triage time, flaky-failure ratio, and escaped-defect rates to capture real-world value.

**Table 1. Comparative View of Modern Testing Tools in AI-Augmented Pipelines**

| Tool | Primary Role | Strengths | Limitations | Best-Fit Context |
|---|---|---|---|---|
| Playwright/Cypress | UI and E2E validation | High debuggability and CI stability | Browser/runtime maintenance overhead | Product UI flows and release gates [11], [16] |
| Testcontainers + Pact/Postman | Integration and contract assurance | Realistic dependency validation | Infrastructure orchestration complexity | Service and API-centric systems [18], [19] |

| k6/Gatling + OpenTelemetry | Performance and observability validation | Deep runtime evidence and capacity insight | Signal interpretation requires mature telemetry | Distributed systems under load [10], [21], [22] |
|---|---|---|---|---|
| CodeQL/Semgrep/OWASP ZAP | Security-aware testing gates | Shift-left risk visibility | False-positive governance needed | Compliance-sensitive delivery [12], [13], [14] |

**Figure 1. Reference Architecture for AI-Driven Testing in CI/CD**

The reference pipeline can be expressed as: requirements and change data -> AI-assisted test generation and prioritization -> deterministic validation gates -> execution across functional, performance, and security layers -> observability-driven diagnosis -> human review for high-impact decisions -> feedback loop to model recalibration. This architecture emphasizes bounded autonomy, auditable decisions, and continuous quality learning [6], [10], [23], [24], [25], [26].

## 5. AI-Driven Test Case Validation

AI-driven validation extends quality assurance beyond static pass/fail checks toward evidence-based confidence estimation. Contemporary systems combine predictive modeling, anomaly detection, runtime observability, and adaptive orchestration to assess both functional correctness and operational reliability. As with generation, the central challenge is scaling automation while preserving trust, explainability, and governance in continuous delivery.

### 5.1 Predictive Models for Test Outcomes

Predictive validation models estimate where failures are likely, which assertions are most informative, and how test outcomes relate to deployment risk. Rather than treating all scenarios equally, these models use historical defect patterns, code-change metadata, component criticality, and prior incident trajectories to prioritize validation depth. The resulting approach improves signal quality by directing computational effort toward areas with higher expected failure impact.

Outcome prediction is typically framed as probabilistic risk scoring. Each change set receives a likelihood profile for regression, performance degradation, or contract breakage, and the pipeline responds by selecting an adaptive validation path. This allows fast paths for low-risk changes while preserving deep verification for high-risk paths such as authentication, payment, and data-consistency workflows.

Anomaly detection adds a complementary layer by identifying deviations that do not match known failure templates. This is important for uncovering novel failure modes, including emergent behavior caused by interaction effects across services. High-maturity systems combine statistical detectors with trace-based context so anomalies can be interpreted causally rather than as isolated outliers.

Risk assessment quality depends heavily on feature fidelity. Effective models incorporate static signals (dependency churn, hotspot density), dynamic signals (latency variance, error-rate spikes), and process signals (release cadence, ownership transitions). This multi-view modeling reduces blind spots and improves calibration of validation priorities over time.

Predictive methods also support selective test-data synthesis. By forecasting which input classes are most likely to expose regressions, models can guide data generation toward boundary and adversarial conditions, increasing defect yield without requiring exhaustive combinatorial expansion. This is especially useful in API and workflow validation where state combinations grow rapidly.

Adaptive validation closes the loop: model predictions are updated continuously from execution outcomes, and subsequent suites are re-ranked accordingly. This online-learning behavior is valuable in fast CI/CD environments, but it requires drift monitoring to prevent stale predictions from silently degrading validation quality.

Performance forecasting is increasingly integrated into validation because many production incidents are non-functional rather than purely logical. Predictive models can estimate latency and throughput risk under expected load profiles, enabling early detection of capacity regressions before release.

From a research perspective, predictive validation should be evaluated using calibration quality, early-defect discovery rate, false-alert burden, and escaped-defect correlation, not only raw classification accuracy. These metrics better reflect operational value in production testing programs.

Overall, predictive validation improves both speed and reliability when integrated with deterministic checks and human oversight. The strongest systems treat predictions as decision support rather than automatic truth, maintaining clear escalation paths for ambiguous or high-impact outcomes.

### 5.2 AI for Continuous Integration/Continuous Deployment (CI/CD)

AI-enhanced CI/CD validation focuses on reducing decision latency while preserving release safety. In practice, this means combining automated execution, risk-aware scheduling, failure clustering, and policy gates so that pipelines can provide rapid yet defensible quality signals. The objective is not simply faster builds, but faster trustworthy decisions.

Automated orchestration remains foundational: tests execute on every relevant change, but AI determines depth, sequence, and scope based on change impact and historical behavior. This approach prevents resource waste on low-value checks while ensuring critical validations are executed early in the pipeline.

Dynamic scheduling and test selection are particularly important in large monorepos and microservice ecosystems, where full-suite execution is often infeasible. AI models can map code deltas to impacted components and prioritize suites with the highest expected defect detection value, reducing cycle time without materially increasing risk.

Self-healing mechanisms improve pipeline stability by repairing non-semantic breakages, such as locator drift or environmental test brittleness. However, robust CI/CD design constrains autonomous repair to low-risk classes and routes behavior-affecting changes through explicit review to avoid silent defect masking.

Predictive failure analysis supports proactive gating by identifying runs likely to fail before full execution completes. Teams can use this signal to trigger early diagnostics, allocate reviewer attention, or run focused confirmatory suites, which reduces triage delay and shortens recovery time.

Optimization in CI/CD also includes intelligent artifact reuse, selective re-execution, and runtime budget control. These methods are critical for scaling quality programs across high-frequency deployments where infrastructure cost and queue time can otherwise erode delivery velocity.

Continuous feedback loops transform CI/CD from a one-way execution path into a learning system. Validation outcomes, flaky signatures, and post-release incidents are fed back into generation and scheduling models, improving prioritization quality across subsequent releases.

Integration with observability and DevOps platforms is essential. AI-driven validation is strongest when test outcomes can be correlated with traces, metrics, and deployment metadata, enabling faster root-cause analysis and clearer release decisions grounded in runtime evidence.

Governed automation is the distinguishing characteristic of mature CI/CD validation. Policy-as-code rules, security checks, and compliance constraints must be enforced alongside functional quality checks so releases are both correct and operationally acceptable.

Therefore, CI/CD effectiveness should be measured through cycle-time reduction, mean time to triage, rollback frequency, and post-deployment defect leakage. These metrics reveal whether automation is truly improving delivery reliability rather than merely increasing pipeline activity.

When implemented with these controls, AI-enhanced CI/CD can substantially improve release cadence while maintaining strong confidence in software quality and operational risk posture.

### 5.3 Self-Healing Test Cases

Self-healing test systems address a persistent operational challenge in automation programs: rapid test decay under frequent UI, API, and environment changes. Their purpose is to preserve suite executability and reduce maintenance burden, but real value depends on controlled adaptation rather than unrestricted automatic modification.

Adaptive script repair generally combines element-level heuristics, semantic context matching, and historical pass/fail patterns to propose corrected interactions when tests break. This improves resilience to superficial interface changes and lowers maintenance effort for repetitive repair tasks.

Element recognition and dynamic locator strategies are now often augmented with visual context and DOM-structure similarity, which improves robustness when identifiers change but behavior intent remains stable. These techniques are especially valuable in web applications with frequent front-end iterations.

Error detection in self-healing pipelines should distinguish environmental instability from genuine functional drift. Effective systems classify breakages by failure signature and only apply autonomous correction when confidence thresholds are met, otherwise escalating to human review.

Context-aware healing is a critical requirement for avoiding false fixes. If a workflow step has changed semantics, preserving old behavior through automatic patching can hide real regressions. For this reason, behavior-affecting changes should trigger validation workflows that include domain-owner approval.

From an engineering-economics perspective, the primary benefit of self-healing is reduced maintenance throughput cost, not elimination of human oversight. Teams can redirect effort from routine repair toward exploratory testing, risk analysis, and architecture-level quality improvements.

Self-healing also supports coverage continuity in long-lived suites by preventing large-scale test attrition after UI refactors or dependency upgrades. Maintaining this continuity improves historical comparability of quality signals across releases.

Learning-from-failure loops further strengthen these systems: each healed or escalated event becomes training evidence that improves future classification and repair quality. Over time, this creates a more stable automation baseline and reduces recurrence of identical break patterns.

Nevertheless, self-healing introduces governance risk if changes are applied opaquely. Mature implementations therefore require auditable change logs, confidence scoring, rollback capability, and approval policies for high-impact workflows.

In summary, self-healing is most valuable as a bounded resilience mechanism inside a broader human-AI quality model. When governed properly, it improves reliability and maintainability; when left unconstrained, it can degrade trust by masking meaningful defects.

## 6. Case Studies and Examples

Real-world adoption patterns show how AI-driven testing performs outside controlled experiments. This section identifies repeatable implementation mechanisms, measurable outcomes, and operational constraints across organizations with different architectures and release profiles. The evidence indicates that durable value emerges when AI is embedded in broader quality workflows rather than treated as isolated automation.

### 6.1 Industry Use Cases

Industry adoption has followed multiple technical paths, yet several convergence trends are visible. Large consumer platforms have used AI-assisted exploration to increase scenario diversity and detect unstable interaction paths in mobile and web ecosystems, while enterprise vendors have integrated AI into broader release-quality programs to improve triage speed and reduce late-cycle defect discovery [1].

Cross-case analysis indicates that organizational context matters as much as model capability. Teams with disciplined CI/CD practices, stable defect taxonomies, and strong service observability generally convert AI recommendations into actionable quality decisions faster than teams lacking those foundations.

At scale, domain criticality shapes governance strategy. High-risk domains such as payments, identity, and compliance-heavy workflows typically apply constrained automation and stronger review gates, while lower-risk surfaces permit broader autonomous assistance. This selective governance pattern is central to sustaining trust in AI-assisted testing.

### 6.2 Impact on Test Efficiency

Efficiency effects are most visible in three dimensions: validation speed, decision accuracy, and fault-relevant coverage. Speed improves through adaptive sequencing and selective execution, which reduces time-to-signal for high-risk changes and enables earlier intervention in the release cycle [1], [1].

Accuracy gains are driven by improved interpretation rather than raw automation volume. AI-supported clustering and anomaly detection help separate true regressions from environmental instability, reducing triage noise and improving confidence in go/no-go decisions. Coverage improves when AI expands boundary and cross-component scenario exploration in ways that manual suites often miss.

### 6.3 Insights from Real World Applications

A key insight from production deployments is that AI functions best as a decision amplifier for expert teams, not as a replacement for engineering judgment. Organizations that retained human ownership of test intent, acceptance boundaries, and risk sign-off achieved more stable long-term outcomes than organizations that treated model output as definitive truth.

Data governance emerged as a decisive enabler. Teams with structured incident taxonomies, reliable telemetry, and clear requirement-to-test traceability produced more actionable model output and lower review overhead. In contrast, weakly labeled or fragmented datasets often produced inflated test volume without proportional defect detection gains [1].

Integration strategy is equally important. Programs that embedded AI outputs into CI/CD gates, service ownership rituals, and post-incident review loops realized compounding quality gains over time, whereas isolated pilots often plateaued after early productivity improvements.

### 6.4 Human Oversight vs Automation

A sustainable human-AI operating model is a control architecture, not a binary preference. AI is highly effective for repetitive execution, large-scale correlation, and prioritization, while human testers remain essential for contextual judgment, ethical interpretation, and resolution of ambiguous failures.

High-maturity teams enforce layered checkpoints: DAI proposes scope and triage hypotheses, and human reviewers validate high-impact decisions before release. This preserves throughput without weakening accountability and reduces the risk of silent quality drift.

The practical objective is therefore not maximum automation, but reliable collaboration between automated inference and expert oversight. Organizations that formalize this balance early typically scale faster and with fewer policy regressions.

### 6.5 Adoption Challenges

Adoption barriers remain significant but manageable through staged rollout. Typical early constraints include legacy compatibility issues, uneven data quality, and limited model operations expertise. Programs that start with bounded pilots, baseline metrics, and clear success criteria generally outperform broad first-wave deployments.
Economic and governance factors are also critical. While AI can reduce long-term maintenance effort, initial investment in integration engineering, tooling, and upskilling is substantial. Sustainable programs combine KPI-driven value tracking with policy controls, auditability, and security-by-design for data-intensive validation pipelines.
Long-term success depends on treating AI testing as an evolving capability: continuous model recalibration, threshold tuning, process refinement, and role clarity are required to preserve quality signal under changing system behavior and business priorities.

**Table 2. Operational Metrics Used to Evaluate AI-Augmented Testing Outcomes**

| Metric | Why It Matters | Typical Improvement Pattern with Governed AI Adoption |
| --- | --- | --- |
| Time-to-signal | Measures early risk detection in pipeline | Faster prioritization and earlier blocker visibility [6], [23], [24] |
| Flaky-failure ratio | Indicates trustworthiness of automated gates | Lower noise after optimization and stability controls [4], [26] |
| Mean time to triage | Captures diagnostic efficiency after failure | Reduced with clustering and observability correlation [10], [25] |
| Escaped-defect rate | Reflects production-side quality leakage | Declines when risk coverage and validation depth improve [1], [23], [28] |
| Release confidence index | Integrates technical and governance evidence | Improves when AI decisions remain auditable and human-validated [6], [27], [29] |

## 7. Discussion

This discussion synthesizes case evidence and examines the balance among benefits, risks, and future directions of AI-driven software testing. It consolidates practical implications for teams that require both delivery speed and trustworthy quality governance.

### 7.1 Benefits of AI in Test Generation and Validation

The most immediate benefit of AI-enabled testing is faster quality feedback under high release velocity. Risk-based prioritization and adaptive execution reduce time-to-signal for high-impact regressions, allowing teams to intervene earlier in the delivery cycle and lower release decision latency [6].

A second benefit is improved coverage quality. AI-assisted generation and optimization can explore boundary and interaction-heavy scenarios that are often underrepresented in manually curated suites, especially in distributed systems with complex cross-service dependencies.
AI also improves triage efficiency by clustering related failures and surfacing likely root-cause patterns. This reduces cognitive overhead during incident analysis and helps teams distinguish product regressions from infrastructure noise more quickly.
Maintainability benefits are equally important. Self-healing and optimization pipelines reduce repetitive script repair effort, enabling quality engineers to focus on exploratory testing, risk modeling, and architecture-level defect prevention.
When integrated with observability, AI-driven testing improves diagnostic depth by linking failed assertions to trace, metric, and log context. This strengthens confidence in failure interpretation and shortens mean time to resolution during release validation.
At the organizational level, AI assistance supports better resource allocation. Validation effort can be dynamically aligned to change risk and business criticality, avoiding uniform full-suite execution while preserving confidence for high-risk paths [1], [1].
These gains compound when measurement discipline is strong. Teams that track early-defect discovery, flaky-failure ratio, triage time, and escaped-defect trends generally extract greater value from AI-enabled workflows than teams that optimize only raw automation volume.
Accordingly, the core benefit of AI in testing is not mere execution scale, but higher-quality decision support. In mature settings, this translates into faster releases, stronger reliability outcomes, and more predictable quality governance.
These benefits are strongest when automation is paired with explicit governance and human accountability, ensuring that increased speed does not come at the expense of interpretability or safety.
From a research perspective, the evidence suggests that AI creates the most value as a quality-intelligence multiplier embedded in disciplined engineering systems rather than as standalone automation.

In practical terms, organizations that operationalize this model report higher release confidence alongside measurable efficiency gains across the testing lifecycle.

### 7.2 Limitations and Risks

Adoption challenges remain substantial, but case evidence indicates they are tractable through staged implementation and governance-first rollout strategies. Early barriers typically include legacy compatibility constraints, fragmented quality data, and limited in-house expertise for model operations [2], [6].
Successful teams usually begin with bounded pilots in high-value workflows, establish baseline metrics, and expand scope only after demonstrating measurable gains in triage speed, defect yield, and release confidence. This incremental path reduces organizational resistance and prevents overcommitment to immature practices.
Economic considerations are also significant. Although AI can reduce long-term maintenance effort, upfront investment in infrastructure, integration engineering, and skill development is non-trivial. Programs that track both technical and business KPIs are better positioned to validate return on investment and secure sustained stakeholder support.
Governance maturity is a central differentiator. Organizations that institutionalize policy checks, audit trails, and model-performance monitoring are more successful at controlling bias, explainability gaps, and compliance exposure. Without these controls, scaling tends to amplify risk faster than value [2], [4].
Vendor and platform decisions also affect adoption trajectory. Selection should prioritize interoperability, transparency, and operational support rather than feature breadth alone, because long-term maintainability depends on how well tools integrate into existing delivery and observability ecosystems.
Change management is often underestimated. Teams need role-specific training, updated review protocols, and shared definitions of model confidence to use AI outputs consistently. Where this social infrastructure is missing, adoption can stall even when technical tools are sound.

Security and privacy requirements add further complexity as AI pipelines increasingly consume production-like data and telemetry. Programs need explicit controls for data minimization, redaction, access governance, and retention to avoid regulatory and operational exposure [12], [13], [14].

Scalability must be planned as an architectural concern from the outset. As adoption expands across services and teams, inference throughput, queueing behavior, and storage growth can become bottlenecks unless capacity and orchestration strategies are proactively designed.

Finally, long-term sustainability depends on treating AI testing as an evolving organizational capability. Continuous retraining, metric review, threshold recalibration, and process refinement are required to maintain quality under changing systems and business priorities.
The strongest adoption pattern is therefore disciplined gradualism: small validated steps, explicit governance, and continuous learning. This pattern consistently outperforms abrupt large-scale rollouts in both quality outcomes and organizational acceptance.
From a research perspective, these adoption dynamics highlight the need for evaluation frameworks that include socio-technical readiness indicators alongside technical performance metrics and realistic software-engineering benchmarks [8], [9].

### 7.3 Future Directions

The strategic implication is clear: adoption success depends less on selecting a single "best" model and more on building a coherent operating system for quality engineering.
When such an operating system is in place, AI functions as a compounding capability that improves over time with each release cycle.
Without that system, even strong models tend to produce intermittent gains that are difficult to sustain.
Therefore, the adoption challenge is ultimately one of institutional design as much as technical implementation.
This conclusion reinforces the broader argument of the paper: progress in modern software testing depends on integrated, governed, and evidence-driven AI use [3], [6].

### 7.4 Threats to Validity

Internal validity is constrained by heterogeneity in organizational maturity, test-data quality, and CI/CD governance across reported case contexts. Observed efficiency gains may partly reflect process improvements that co-occur with AI adoption rather than model effects alone. To mitigate this risk, the analysis compares outcomes across multiple technical patterns and explicitly separates capability claims from deployment prerequisites.

External validity is limited by domain effects. Evidence from web and enterprise service ecosystems may not fully generalize to safety-critical, embedded, or highly regulated environments where verification obligations and assurance thresholds differ materially. Construct validity is also a concern because common metrics, such as pass rate and coverage, can overstate real defect-prevention value if not combined with mutation sensitivity, escaped-defect trends, and triage-quality indicators [15], [26].

Finally, temporal validity should be considered: AI testing tooling and model behavior evolve quickly. Findings grounded in a 2024-2026 evidence window require periodic re-evaluation as model capabilities, benchmark standards, and governance expectations continue to shift. Future longitudinal studies should therefore report recalibration intervals and policy-drift effects explicitly [23], [27], [28], [29].

## 8. Conclusion

This paper has shown that AI-driven software testing is most effective when implemented as a governed quality engineering system rather than as isolated automation. Across generation, optimization, and validation stages, the strongest outcomes come from combining probabilistic AI capabilities with deterministic controls, observability evidence, and explicit human accountability. In this integrated model, AI improves decision speed, expands fault-relevant coverage, and strengthens triage quality, while engineers retain authority over risk interpretation and release acceptance.

The analysis also highlights that long-term value depends on socio-technical maturity: reliable data pipelines, transparent model behavior, policy-aware CI/CD integration, and continuous measurement against operational quality metrics. Under these conditions, organizations can reduce escaped defects, lower maintenance burden, and increase release confidence without sacrificing governance. Future progress should therefore focus on trustworthy autonomy, where AI systems remain adaptive and efficient while preserving auditability, explainability, and alignment with organizational risk tolerance.